# Photoionization rates of oxygen and nitrogen at 800 nm measured by Thomson coherent microwave scattering

K. Reindersma, M. Malaussena, S. Anantaram, K. Patel, V. Kumar, N. Babusis, M. Slipchenko, A. Shashurin
*Purdue University, West Lafayette, IN 47907, USA*

**In this study, we report on the use of Thomson coherent microwave scattering (CMS) to characterize photoionization rates in oxygen and nitrogen at 800 nm over a range spanning approximately three orders of magnitude, from about mid-$10^9$ $s^{-1}$ range to the to mid-$10^{12}$ $s^{-1}$ range, corresponding to laser intensities in the vicinity of $10^{13}$-$10^{14}$ W/cm$^2$, a range of significant practical relevance to filamentation physics.**

## I. Nomenclature

| | | |
|---|---|---|
| *CMS* | = | Coherent Microwave Scattering |
| *HHG* | = | High-Order Harmonic Generation |
| *FWHM* | = | Full-Width Half-Maximum |
| *PTFE* | = | Polytetrafluoroethylene |
| *REMPI* | = | Resonant Enhanced Multiphoton Ionization |

## II. Introduction

Photoionization (PI) rates are fundamental physical constants that are largely unknown. Knowledge of photoionization rates of air (and its constituents) is critical in the fields of nonlinear optics and filamentation physics[1,2,3] and corresponding applications including lightning protection,[4] atmospheric remote sensing,[5,6] generation of terahertz and gigahertz emission,[7,8,9] laser controlled weather phenomena,[10,11] microwave guiding,[12,13] high-order harmonic generation (HHG),[3,14] etc. Photoionization rate is the single most important parameter (for both numerical simulations and experiments) to control the electron number density for achieving the specific purposes of the aforementioned applications. By far, the most relevant wavelength for filamentation physics is 800 nm due to its wide availability through femtosecond Ti:sapphire lasers and low absorption in air. Other wavelengths previously utilized for laser filamentation in the atmospheric air span from the UV range at 267 and 400 nm (third and second harmonics of Ti:sapphire laser systems) to the mid-IR at 3.9 μm.[1,2,4,15,16,17,18,19] Knowledge of photoionization rates is also crucial in advancing multiple optical diagnostic techniques utilized in the fields of gaseous discharges, combustion, and high-speed aerodynamics, such as electric-field induced second harmonic generation (E-FISH),[20,21,22,23,24,25,26,27,28,29,30,31,32,33,34] two-photon laser induced fluorescence (TPLIF),[35,36,37,38,39,40,41,42,43] and femtosecond laser electronic excitation tagging (FLEET) velocimetry. [44,45,46,47,48,49]

Current limitations to knowledge of PI rates are largely due to challenges associated with electron number density diagnostics, as sizes of photoionized plasma volumes are typically small (sub-millimeter). Therefore, even though the dependence of ionization rate on laser intensity and wavelength for atomic gases is theoretically described in seminal works,[50,51,52] the empirical data on photoionization rates is lacking. A number of semi-empirical methods for relative measurements of plasma density were proposed for near-IR laser-induced plasma, including ion current measurement by time-of-flight mass spectrometer, scattering of THz radiation, and measurements of capacitive response times; however, these require absolute calibration based upon theoretically predicted values of plasma number density.[53,54,55,56,57,58,59,60,61] Capacitive probes were recently applied to estimate plasma density and electric conductivity in mid-IR filaments,[53,61] but this approach is controversial due to uncertainty in the interpretation of the positive current swing used for determining electron numbers in the plasma volume. Photoionization rates inferred from the direct counting of photoionized electrons based on picosecond avalanche amplification of individual electrons was reported; however, these rates were not determined entirely experimentally (in the intensity range most relevant to filamentation physics) and instead were anchored to a theoretical ionization yield at upper intensity endpoint (100 TW/cm$^2$), with the initial electron density inferred from avalanche-breakdown timing under the assumption of a constant electron multiplication rate. Consequently, these results do not constitute a fully theory-independent determination of the absolute ionization yield at intensities in the range 10-100 TW/cm$^2$ (a range relevant to laser filamentation).[62]

In our recent work, we developed a methodology for measuring PI rates based on a coherent microwave scattering (CMS) approach and demonstrated the feasibility and potential of the method. Specifically, PI rates of oxygen in near-IR (at 800 nm) and PI rates of air in mid-IR (at 3.9 um) were measured using CMS. [63,64]

However, these recent reports were conducted at atmospheric pressure and, correspondingly, in the collisional CMS regime, which introduces several limitations. The primary constraint is that operation at atmospheric pressure limits the maximum usable laser intensity because nonlinear optical effects associated with refractive-index modification, including Kerr and plasma nonlinearities, arise at relatively low intensities; e.g., an upper intensity limit of approximately 7 TW/cm$^2$ was reported for collisional CMS measurements of oxygen photoionization rates at 800 nm.[63] Additionally, collisional CMS requires knowledge of electron-gas collisional frequency which cannot be precisely evaluated and must be approximately estimated instead (via assumptions on electron energy distribution). This introduces additional uncertainty and, consequently, reduces the accuracy of the determined photoionization rates.

Due to these limitations of collisional CMS at atmospheric pressure, our subsequent work investigated and successfully demonstrated the feasibility of operating CMS at low pressures in the Thomson regime ($p \leq 1$ Torr for probing CMS frequency ~10 GHz) where the aforementioned deficiencies are mitigated.[65,66,67] The Thomson regime offers several important advantages over collisional CMS: (i) in the Thomson regime, when the electron-gas collision frequency is small compared to the probing microwave frequency ($\nu_m \ll \omega$), precise knowledge of $\nu_m$ is unnecessary to determine the total electron count in plasma volume ($N_e$) from the scattered signal, in contrast with the collisional regime; (ii) operation at low pressures ($p \leq 1$ Torr) enables measurements at higher laser intensities due to the delayed onset of nonlinear optical phenomena.[65,66,67] This refers to the fact that nonlinear optical phenomena including Kerr and plasma nonlinear terms in the refractive index reduce proportionally with a decrease in pressure; (iii) operation at low pressure slows the electron decay, allowing CMS to accurately measure $N_e$ by avoiding underestimation associated with rapid, sub-nanosecond electron decay at atmospheric pressure.

Therefore, these recent studies open the possibility of using Thomson-regime coherent microwave scattering (Thomson CMS) for precise empirical determination and tabulation of photoionization rates for processes of high practical importance. However, no photoionization rates have yet been determined using this approach, and only limited work with Thomson CMS at low pressures has been conducted to date. Specifically, prior studies have demonstrated the feasibility of Thomson CMS through measurements of $N_e$ in laser-induced plasmas generated by resonance-enhanced multiphoton ionization (REMPI) of oxygen and krypton.[65,66,67] In this work we used Thomson CMS to measure photoionization rates in oxygen and nitrogen at 800 nm in a range of significant practical relevance to filamentation physics.

## III. Methodology

A custom vacuum chamber was constructed for conducting measurements of the photoionization rates in various gases and pressures, as shown in Figure 1. A 24” cubic stainless steel vacuum chamber with walls covered with microwave absorber panels was utilized to provide an anechoic environment ideal for isolating the scattered signal. The chamber was pumped by a diffusion pump and uses fused silica viewports for laser access, a microwave instrumentation feedthrough, and additional ports for process gas handling and auxiliary instrumentation. The chamber was filled with various gas types at controlled pressure level. The vacuum testing facility was relatively compact and mobile, allowing an easy transfer between the labs to access various laser facilities.

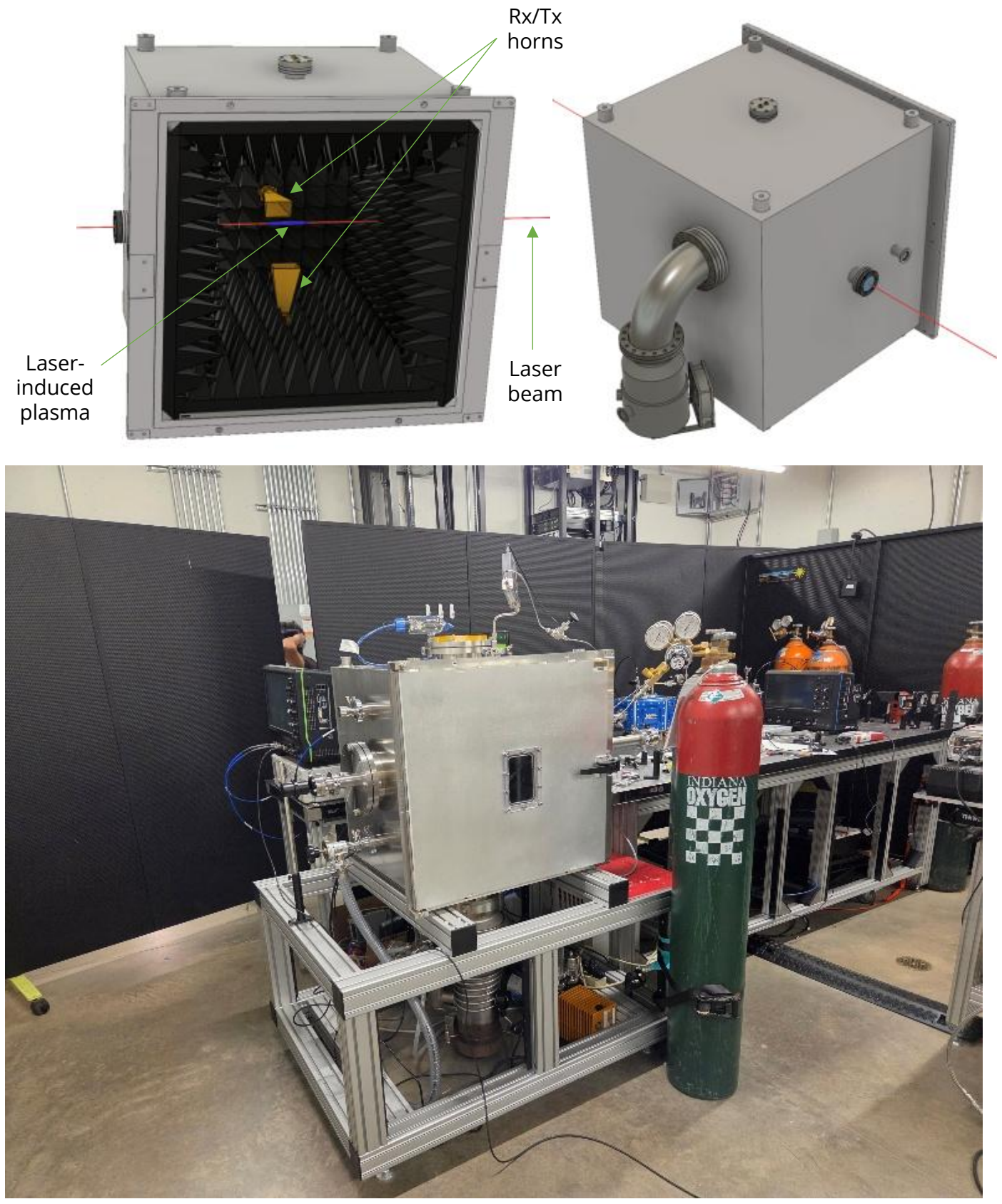

**Figure 1. A 24” cubic stainless steel vacuum facility constructed for measurements of photoionization rates using coherent microwave scattering in Thomson regime.**

The overall experimental system utilized in this work is illustrated in Figure 2. The system utilized an 800 nm wavelength Ti:Sapphire femtosecond laser and a Thomson CMS system to probe the laser-induced plasma at low pressures inside a cubical vacuum chamber. The ambient temperature was assumed to be 294 K during operation. The repetition rate of 10 Hz was utilized in this work to ensure complete plasma decay between successive laser pulses and thereby entirely eliminate memory effects. The laser beam was focused using a plano-convex 1000 mm focal length lens. The beam diameter prior to focusing was measured as 10.8 mm from the beam profiling camera, having approximately centered the laser at a corner of the sensor and the dimensions of the sensor being 7.7 mm x 5.4 mm. The pulse energy of the laser was taken from 0.6 to 2.2 mJ at the maximum and was measured using a laser power meter (Gentec-EO XLP12-3S-H2-DO). The laser induced plasma was visually observed using an ICCD camera (Princeton Instruments PI-MAX 1024i) and a regular camera.

The total number of electrons was obtained through the use of a Thomson CMS system technique. The system used a homodyne detection circuit to measure the amplitude and phase of the microwave signal scattered by the plasma, as illustrated in Figure 2. Specifically, an Anritsu 68369B synthesized signal generator (10.5-12 GHz range, FWHM 1 MHz) was used to produce the microwave signal, which is then amplified and split into two arms. One arm was connected to a HMC-042 I/Q mixer local oscillator (LO) port, with a confirmed operational power in the linear mixer regime (18-19 dBm). The other arm was then further amplified, isolated, and sent to a pyramidal 20 dB transmitting (Tx) horn for plasma irradiation. Negligible heating of the plasma by probing microwaves was confirmed though the invariance of plasma decay behavior on modest increases in supplied microwave power (0.1 W→6 W). A 15 dB receiving horn was then installed and connected to the I/Q mixer radio frequency (RF) port to detect the scattered electric field. Both the transmitting and receiving horns were placed inside the vacuum chamber at a proper far-field distance of ~20 cm from the plasma ellipsoid center - where sealed SMA feedthroughs are installed for external connections to the I/Q mixer. The in-phase and quadrature channels of the mixer were subsequently amplified (Pasternack PE15A1007 - 32dB gain, 9 kHz – 3 GHz bandwidth) and connected to a Teledyne LeCroy HDO-9304 3.5 GHz oscilloscope with 50 Ohm DC termination. Calibrated amplification of the microwave signals in the circuit was used as needed to improve system sensitivity, always ensuring that all amplifiers operate in an undistorted linear regime throughout experiments. The absolute calibration of the CMS was conducted using a dielectric scattering PTFE cylinder shot through the probing region of the microwave antennas.

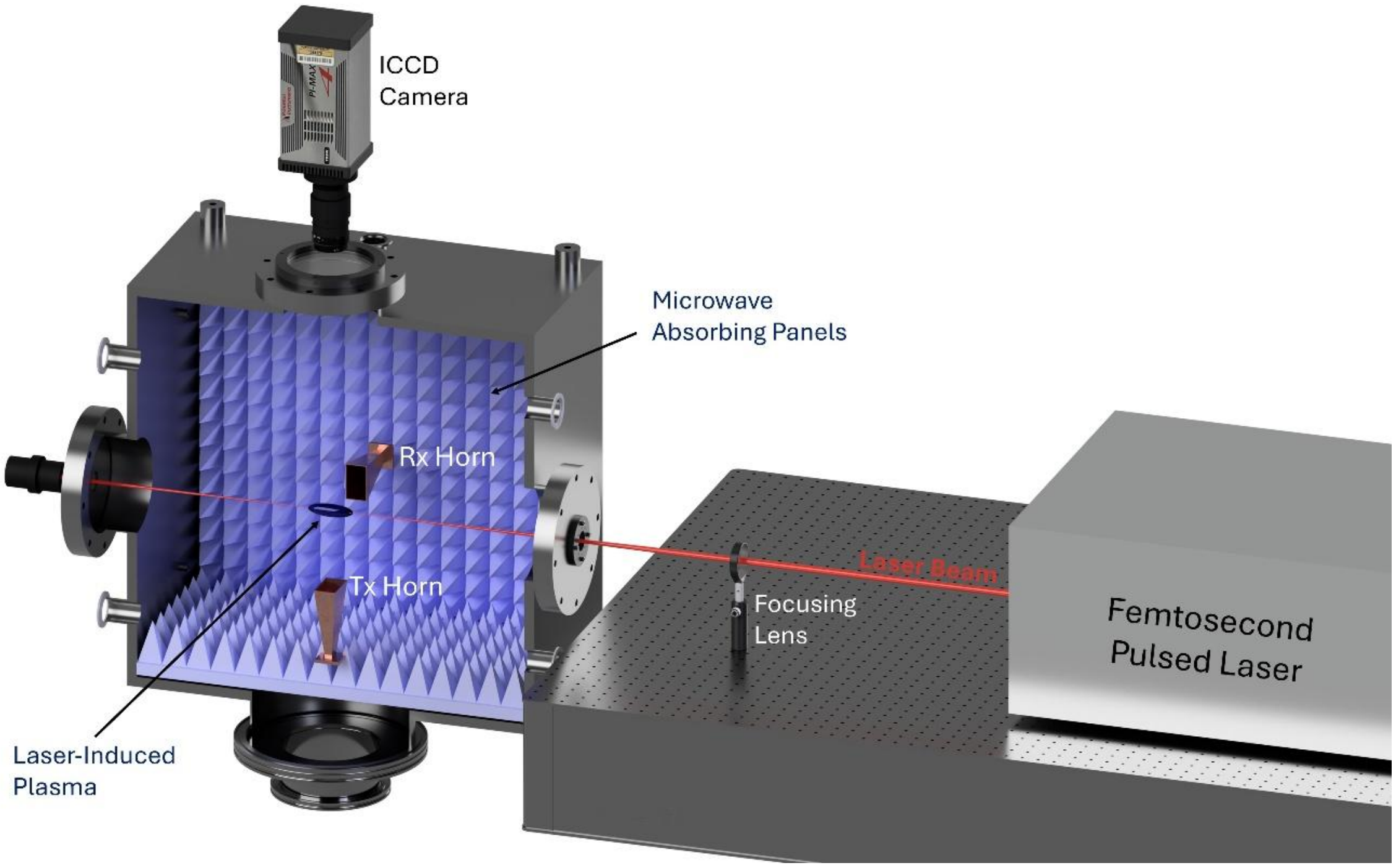

**Figure 2. Experimental system used in this work, including an 800 nm wavelength Ti:Sapphire femtosecond laser and a Thomson CMS system to probe the laser-induced plasma inside a cubical vacuum chamber.**

## IV. Results & Discussion

Photoionization rate was determined from the measured $N_e$ using the following procedure. The procedure assumes that the dependence of $\nu_{PI}$ on laser intensity can be locally approximated by a power law, $\nu_{PI} \propto I^{\alpha}$ , within the vicinity of a specific intensity range of interest (The validity of this assumption is subsequently verified using the experimentally determined $\nu_{PI}$ dependence on intensity). Under this assumption, the plasma formation due to the laser pulse is governed by the simple differential equation $\frac{\partial n_e}{\partial t} = \nu_{PI} n_g$, assuming a low ionization degree ($n_e \ll n_g$) and that the plasma electrons remain effectively collisionless over the duration of the laser pulse ($\nu_m \tau_{FWHM} \ll 1$). Here, $n_e$ is the electron number density, $n_g$ is the background gas number density, and $\tau_{FWHM}$ is the FWHM pulse duration. In this case, the total number of electrons generated by the laser pulse, $N_e = n_g \iint \nu_{PI} dt dV$, within the laser-induced plasma volume, can then be analytically integrated and reduced to the form $N_e = \frac{\pi^2}{2\sqrt{\alpha \ln 2}} \cdot \frac{1}{2\alpha} \cdot \frac{\Gamma\left(\alpha - \frac{3}{2}\right)}{\Gamma(\alpha - 1)} \cdot n_g \nu_{PI} \tau_{FWHM} \cdot w_0^2 \cdot z_R$ , where $\nu_{PI}$ is the photoionization rate in center of laser-induced plasma volume, corresponding to the peak intensity in the center $I_0 = \frac{E_L}{\frac{1}{2}\pi w_0^2} \frac{2\sqrt{\ln 2}}{\tau_{FWHM}\sqrt{\pi}}$, $w_0$ is the beam waist radius, and $z_R$ is the Rayleigh length. Finally, the photoionization rate corresponding to the intensity $I_0$ is determined as:

$$\nu_{PI} = \frac{N_e}{\frac{\pi^2}{2\sqrt{\alpha \ln 2}} \cdot \frac{1}{2\alpha} \cdot \frac{\Gamma\left(\alpha - \frac{3}{2}\right)}{\Gamma(\alpha - 1)} n_g w_0^2 z_R \tau_{FWHM}} \qquad \text{Eq. (1)}$$

The laser-pulse characteristics ($w_0, z_R$, and $\tau_{FWHM}$) were measured as described below in this section.[63,65,68] The power-law exponent $\alpha$ was determined from the slope of the experimentally measured $N_e$ vs. $I_0$ plotted in log-log scale. Note that $\alpha$ was observed to gradually decrease with increasing intensity, corresponding to the transition of the photoionization regime from multiphoton ionization to tunneling ionization.

Spatiotemporal characterization of the focused laser beam was conducted using a strongly attenuated beam, with several optical-density filters used to prevent damage to the beam-profiling camera. The beam profiler was mounted on an optical translation stage and scanned along the beam-propagation direction to measure the beam radius in the vicinity of the beam waist, as shown in Figure 3 (a). The measured beam radius evolution was fitted using Gaussian-beam propagation, yielding a beam-waist radius of $w_0$= 80.41 µm and Rayleigh length $z_R$= 17.5 mm. Temporal characterization was conducted by measuring the intensity autocorrelation function using a second-harmonic-generation crystal, as shown in Figure 3 (b). The temporal full width at half maximum (FWHM) of the laser pulse was determined to be 102.91 fs. A typical image of the laser-induced plasma is shown in Figure 3(c).

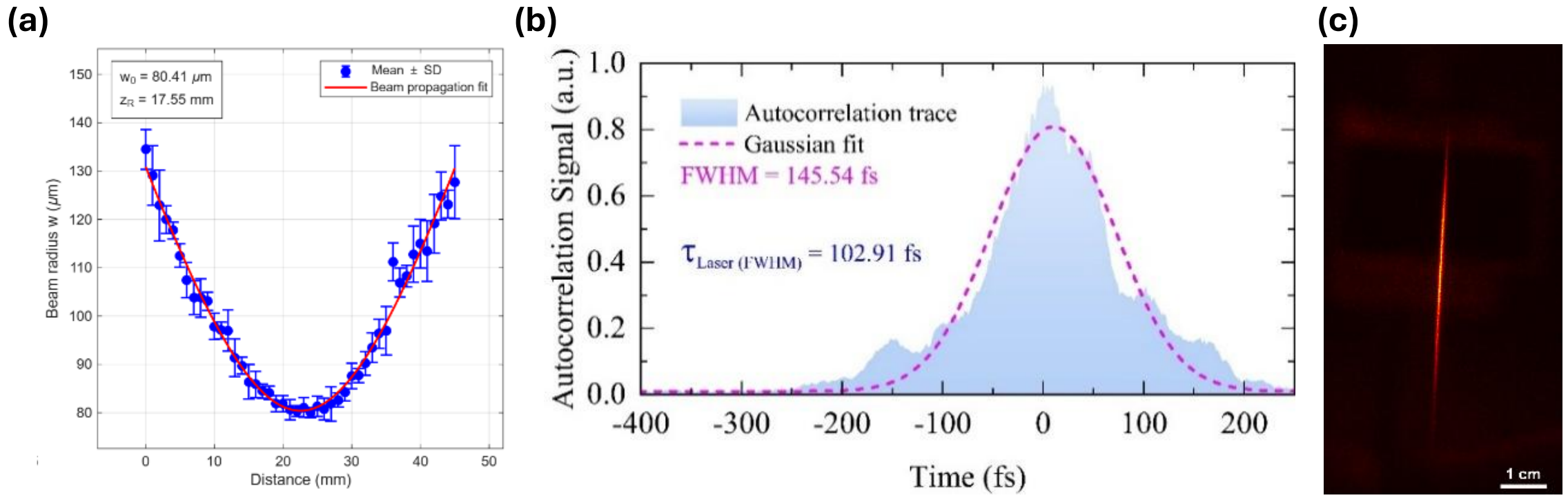

**Figure 3. (a) Measured and fitted beam radius in the vicinity of the beam waist. (b) Temporal characterization of the laser pulse based on measurements of the intensity autocorrelation function. (c) Typical image of the laser-induced plasma.**

Figure 4 shows typical temporal dynamics of the measured scattered signal for oxygen and nitrogen at laser pulse energies of 1.51 and 1.92 mJ, respectively. All experiments were conducted at a pressure of 1 Torr of oxygen or nitrogen, and the absence of nonlinear optical effects (including Kerr and plasma nonlinearities) was confirmed by observing no measurable shift in the location of the laser-induced plasma volume relative to its position at lower pulse energies, where nonlinear effects are known to be negligible.[68] Two distinct processes can be observed in Figure 4: a rapid rise in the scattered signal associated with ionization of the test gas by the femtosecond laser pulse, followed by a much slower sub-microsecond decay. This decay time is significantly longer than that previously reported at atmospheric pressure due to the reduced gas pressure in the chamber.[63,68] Although the CMS system does not temporally resolve the photoionization process itself, the peak value of the scattered signal corresponds to the total number of electrons generated by the laser pulse. This peak value was therefore used to determine the total number of electrons $N_e$ produced by the laser pulse.

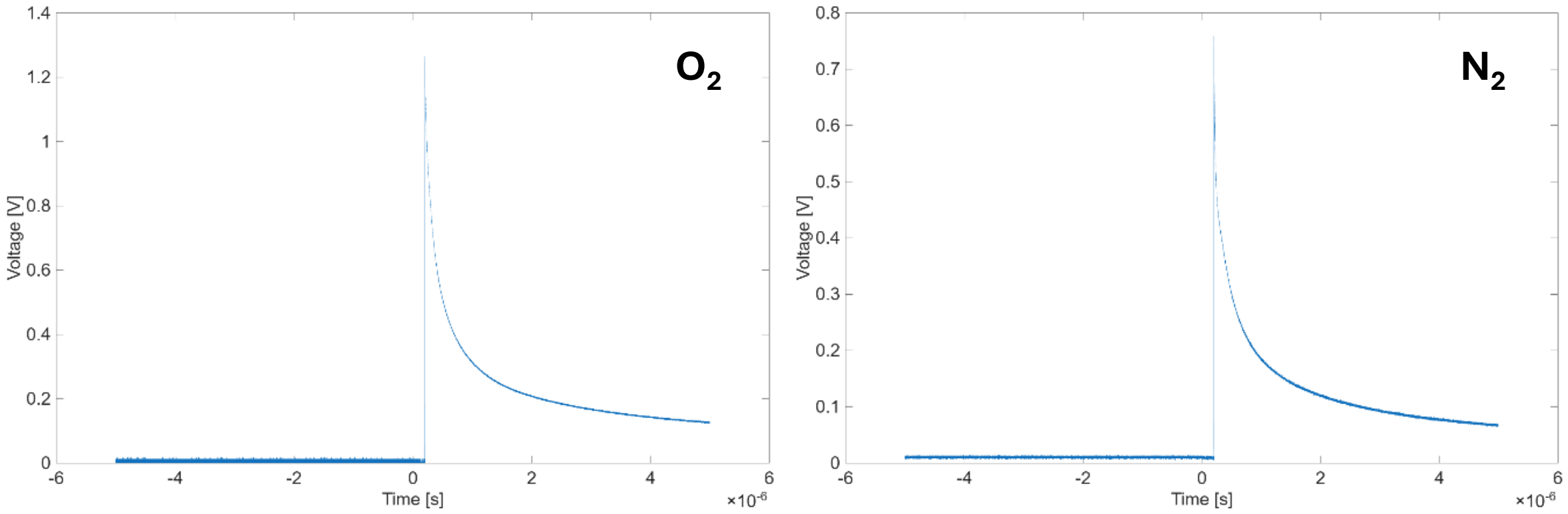

**Figure 4. Typical temporal dynamics of the measured scattered signal for oxygen and nitrogen (laser pulse energies of 1.51 and 1.92 mJ, respectively, are shown).**

Finally, Figure 5 shows the dependence of the photoionization rate, $\nu_{PI}$, on intensity, obtained using the methodology described above from the experimentally determined $N_e$, $\alpha$, $w_0, z_R$, and $\tau_{FWHM}$ using Eq. (1). One can see that the measurements span approximately three orders of magnitude in $\nu_{PI}$, from about mid-$10^9$ $s^{-1}$ range to the to mid-$10^{12}$ $s^{-1}$ range, corresponding to laser intensities in vicinity of $10^{13}$-$10^{14}$ W/cm$^2$. This measurement range is particularly relevant to filamentation physics.

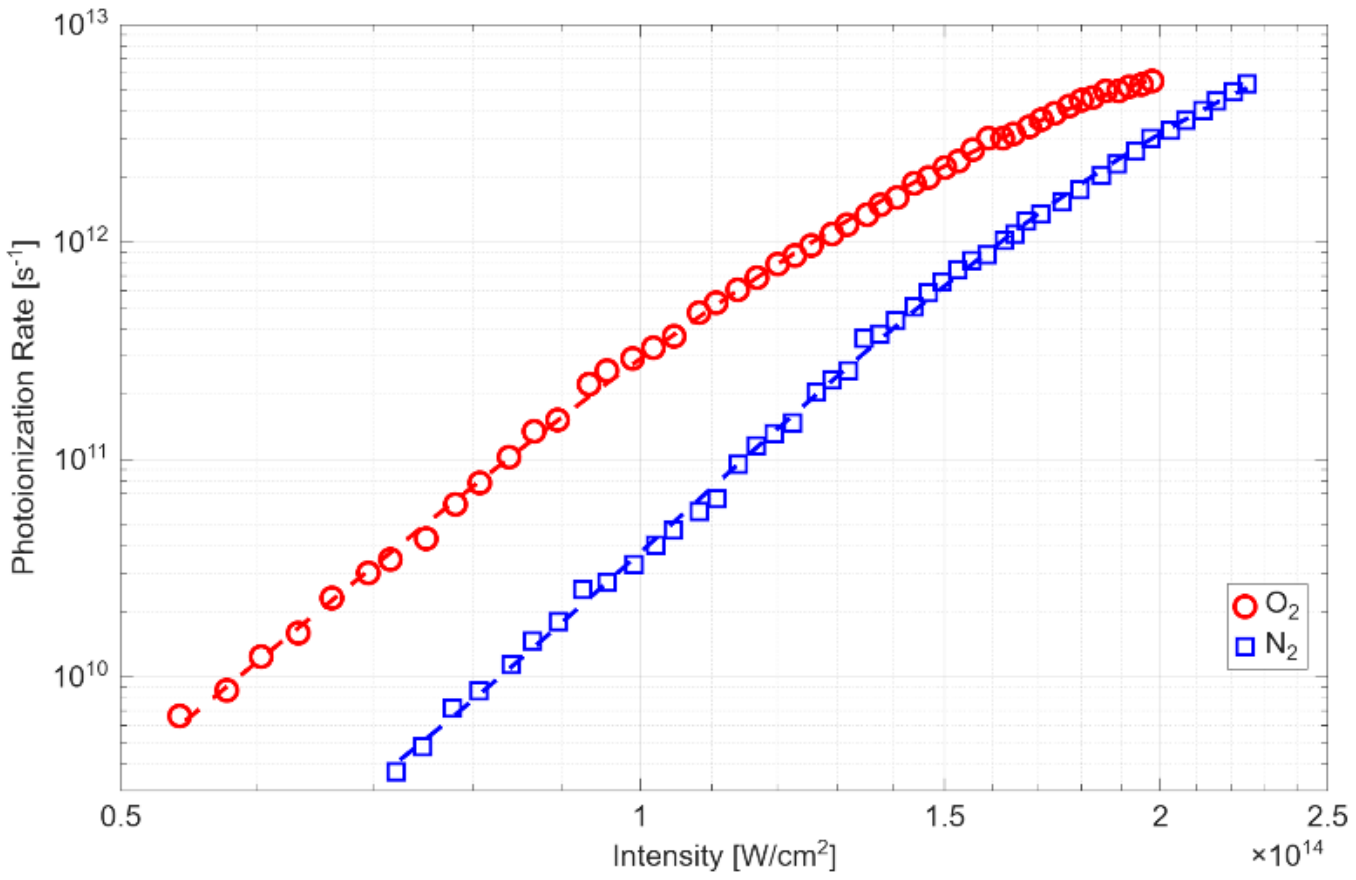

**Figure 5. Dependence of the photoionization rate on intensity for oxygen and nitrogen at 800 nm spanning approximately three orders of magnitude in $\nu_{PI}$ and corresponding to laser intensities in the vicinity of $10^{13}$-$10^{14}$ W/cm$^2$.**

## V. Conclusion

This work demonstrates the feasibility of using the coherent microwave scattering technique for absolute measurements of the photoionization rates using the Thomson scattering regime. Using this method, photoionization rates for oxygen and nitrogen at 800 nm were quantified across approximately three orders of magnitude, extending from the mid -$10^9$ $s^{-1}$ range to the to mid-$10^{12}$ $s^{-1}$ range, and corresponding to laser intensities in the vicinity of $10^{13}$-$10^{14}$ W/cm$^2$ range. These measurements cover a range of particular practical importance for filamentation physics.

This work was supported by the U.S. National Science Foundation (Grant No. 2409559).